%% file: iclp2001.tex
\documentclass{llncs}
\usepackage{latexsym}
\usepackage[dvips]{graphicx}
\usepackage{amsmath}

\begin{document}
\author{Nancy Mazur\inst{1} \and 
	Peter Ross\inst{2}
	\and 
	Gerda Janssens\inst{1} \and 
	Maurice Bruynooghe\inst{1}}
\authorrunning{Nancy Mazur et al.}
\institute{
Department of Computer Science, K.U.Leuven \\
Celestijnenlaan, 200A, B--3001 Heverlee, Belgium \\
\email{\{nancy,gerda,maurice\}@cs.kuleuven.ac.be}\\
\and
Mission Critical, Dr\`eve Richelle, 161, B\^at.\ N \\
B--1410 Waterloo, Belgium \\
\email{petdr@miscrit.be}}
\title{Practical Aspects for a Working Compile Time\\
Garbage Collection System
for Mercury}
\titlerunning{Memory reuse in Mercury}
\maketitle

\begin{abstract}
Compile-time garbage collection (CTGC) 
is still a very uncommon feature within
compilers. 
In previous work we have developed a compile-time structure
reuse system for Mercury, a logic programming language. 
This system 
indicates which datastructures can safely be reused at run-time. 
As preliminary experiments were promising, we 
have continued this work and have now a 
working and well performing near-to-ship
CTGC-system built into the Melbourne Mercury
Compiler (MMC). 

In this paper we present the multiple design decisions leading to this
system, we report the results of using CTGC 
for a set of benchmarks, including a
real-world program, and finally we discuss further
possible improvements. 
Benchmarks show substantial memory savings and a noticeable reduction
in execution time. 
\end{abstract}


\section{Introduction}

Modern programming languages typically limit the possibilities of 
the programmer to manage memory directly. In such cases allocation and
deallocation is delegated to the run-time system and its garbage
collector, at the expense of possible run-time overhead. 
Declarative languages go even further by prohibiting 
destructive updates. This increases the run-time overhead
considerably: new datastructures are created instead of updating
existing ones, hence garbage collection will be needed more often. 

Special techniques have been developed to overcome this handicap and to 
improve the memory usage, both for logic programming 
languages~\cite{Gudjonsson:1999:CTM,kluzniak.gc,anne.toplas94}
and functional
languages~\cite{total97-for-dart,Mohnen:DISS:97}. 
Some of the approaches depend on a combination of special language
constructs and
analyses using unique objects~\cite{mercury.jlp96,ilps95*51,wadler:92},
some are solely based on compiler
analyses~\cite{iclp97*18,Mohnen:DISS:97}, and 
others combine it with special memory layout
techniques~\cite{total97-for-dart}. In this work we develop a purely
analysis based memory management system. 

Mercury, a modern logic programming language with 
declarations~\cite{mercury.jlp96} profiles itself as a general 
purpose programming language for large industrial projects. Memory
requirements are therefore high. Hence we believe it is a useful
research goal to develop a CTGC-system for this language. 
In addition, mastering it for Mercury should
be a useful stepping stone for systems such as Ciao Prolog~\cite{Ciao99}
(which has optional declarations and includes the impurities of
Prolog) and HAL~\cite{HAL99} (a Mercury-based constraint language).

The intention of the CTGC-system is to discover at compile-time
when data is not referenced anymore, and how it can best be reused. 
Mulkers et al.\ \cite{anne.toplas94} have developed an analysis for
Prolog which detects when memory cells become available for reuse. 
This analysis was first adapted to languages with 
declarations~\cite{ICLP97} and then refined for use in the presence
of program modules~\cite{31661}. 
A first prototype implementation was made to measure the potential of 
the analysis for detecting dead memory cells. 
As the results of the prototype were promising~\cite{31661}, 
we have continued this
work and implemented a full CTGC-system for the Melbourne Mercury Compiler
(MMC), focusing on minimizing the memory 
usage of a program.
In this paper we present the different design decisions
that had to be taken to obtain noticeable memory savings,
while remaining easy to implement within the MMC and with acceptable
compilation overhead.
A series of benchmarks are given,
measuring not only the global effect of CTGC,
but also the effect of the different decisions during the CTGC analysis.




\vspace{0.2cm}
After presenting some background in
Section~\ref{sec:background}, we first solve the problem of deciding
how to perform reuse once it is known which cells might die
(Section~\ref{sec:selection}).
Section~\ref{sec:additions} presents low-level additions
required to increase precision and speed, 
and obtain the first acceptable results for a set of benchmarks
(Section~\ref{sec:results}). Using cell-caching
(Section~\ref{sec:cellcache}) more memory savings can 
be obtained.
Finally improvements related to other work are
suggested (Section~\ref{sec:improvements}), followed
by a conclusion (Section~\ref{sec:conclusion}).


\section{Background}
\label{sec:background}

\subsection{Mercury}

Mercury~\cite{mercury.ref} is a logic programming 
language with types, modes and
determinism declarations. Its type
system is based on a polymorphic many-sorted logic
and its mode-system does not allow partially
instantiated datastructures.

The analysis performed by our CTGC-system is at the level
of the {\em High Level Data
  Structure} (HLDS) constructed by the MMC. Within this
structure, predicates are {\em normalized}, i.e. all atoms appearing
in the program have distinct variables as arguments, and all
unifications $X = Y$ are explicited as one of 
(1) a test $X == Y$ (both are ground terms), 
(2) an assignment $X := Y$, ($X$ is free, $Y$ is ground)
(3) a construction $X \Leftarrow f(Y_1,\ldots,Y_n)$ ($X$ is free, 
all $Y_i$ are ground), 
or 
(4) a deconstruction $X \Rightarrow f(Y_1,\ldots,Y_n)$
($X$ is ground, all $Y_i$ are free)~\cite{mercury.ref}.
Within the HLDS, the atoms of a clause body
are ordered such that the body is well
moded. In the paper, we will use the explicit modes.

Just like in the HLDS we will use the notion of a {\em procedure}, i.e.
a combination of one predicate with {\em one} mode, and thus talk
about the analysis of a procedure. 

\subsection{General Structure of the CTGC-System}
\label{sec:generalstruct}

The CTGC-system consists of a data-flow
analysis, followed by a reuse analysis and ended by a code generation 
pass (similar to~\cite{Gudjonsson:1999:CTM}).

The data-flow analysis is performed to 
obtain structure-sharing information 
(expressed as possible aliases~\cite{ICLP97})
and to detect when heap
cells become {\em dead} and are therefore {\em available for reuse}.
It is based on abstract interpretation~\cite{mb.jlp} using
a so called {\em default call pattern} for each of the procedures
to be analysed. This default call pattern makes minimal realistic
assumptions: the inputs to a procedure are in no way
aliased, and only the outputs will be used after the call to the
procedure. 
The data-flow analysis requires a fixpoint computation to deal 
with recursive predicate definitions. 
For more details, see~\cite{ICLP97,31661}. 

Next the reuse analysis 
decides which reuses are possible (see Section~\ref{sec:typesreuse}).
Different versions can then be
created for the different reuses detected. 
While the underlying concepts were already developed in~\cite{31661}, the
pragmatics of our implementation are discussed in this paper. 

Finally, low-level code corresponding to the detected reuses is generated.

As Mercury allows programming with modules, the CTGC-system processes
each module independently. Interface files are used to allow 
analysis information (structure-sharing and reuse information) 
generated while processing one module 
to be used when processing other modules.

\subsection{Data Representation}
\label{section:repr}

The purpose of the CTGC-system is to identify which objects on the heap, 
so called {\em datastructures}, become dead and can 
therefore be reused. In order to understand what these objects are, we 
will clarify the way typed terms are 
usually\footnote{The MMC compiles to different back-ends, the most
common being ANSI-C. Higher-level back-ends, such as Java or .NET, 
use different low level representations, yet the theory of 
recycling heap cells remains the same.}
represented in the MMC. 
Consider the following types: 
\begin{verbatim}
    :- type dir ---> north ; south ; east ; west.
    :- type example ---> a(int, dir) ; b(example).
\end{verbatim}
Terms of primitive types such as integers, chars, 
floats\footnote{Depending on the word-size, these might have a 
boxed representation.} and pointers to strings 
are represented as single machine words. 
Terms of types such as {\tt dir}, in which every alternative is a constant are
equivalent to enumerated types in other languages.  
Mercury represents
them as consecutive integers starting from zero, and stores
them in a single machine word. 
Terms of types such as {\tt example} are stored on the heap. 
The pointer 
to the actual term on the heap is tagged~\cite{dowd99run}. 
This tag is used to indicate the function symbol of the term.
Terms of types 
having more function
symbols than a single tag can distinguish use secondary tags.

\begin{figure}[bt]
\begin{minipage}[b]{12cm}
\begin{minipage}[b][4.5cm][c]{6cm}
\begin{center}
\includegraphics[width=4.5cm]{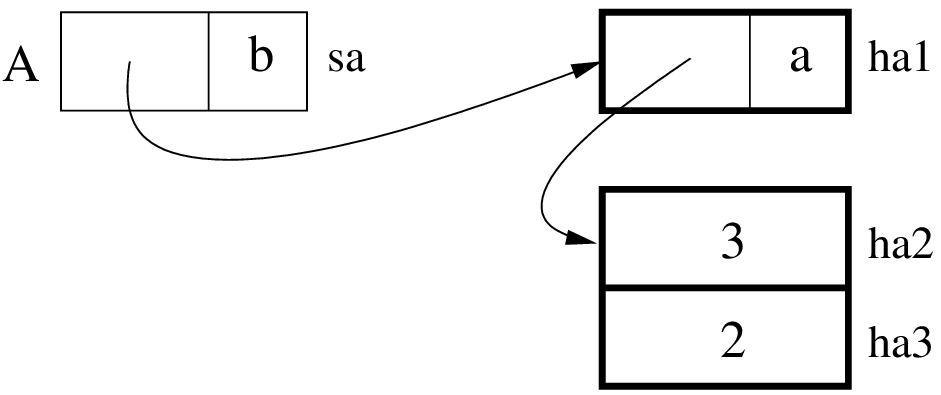}\par
\end{center}
\caption{{\tt A = b(a(3,east))}.}
\label{fig:lowlevelX}
\end{minipage}
\hfill
\begin{minipage}[b][4.5cm][c]{6cm}
\begin{verbatim}
:- pred convert1(example, example).
:- mode convert1(in, out) is semidet.
convert1(X,Y):- X => b(X1), 
               X1 => a(A1, _), 
               Y1 <= a(A1, north), 
               Y <= b(Y1).
\end{verbatim}
\caption{Conversion-procedure.}
\label{fig:convert1}
\end{minipage}
\end{minipage}
\end{figure}

Figure~\ref{fig:lowlevelX} shows the representation of 
a variable {\tt A} bound to {\tt b(a(3,east))}. 
In this paper
{\tt ha1}, {\tt hy1},\ldots denote heap cells, whereas
{\tt sa}, {\tt sx},\ldots  are registers or stack locations.

\subsection{Data Reuse}
\label{section:reuse}

\begin{figure}[tb]
\begin{minipage}[b]{12cm}
\begin{minipage}[b][4.5cm][c]{6cm}
\begin{center}
\includegraphics[width=4.5cm]{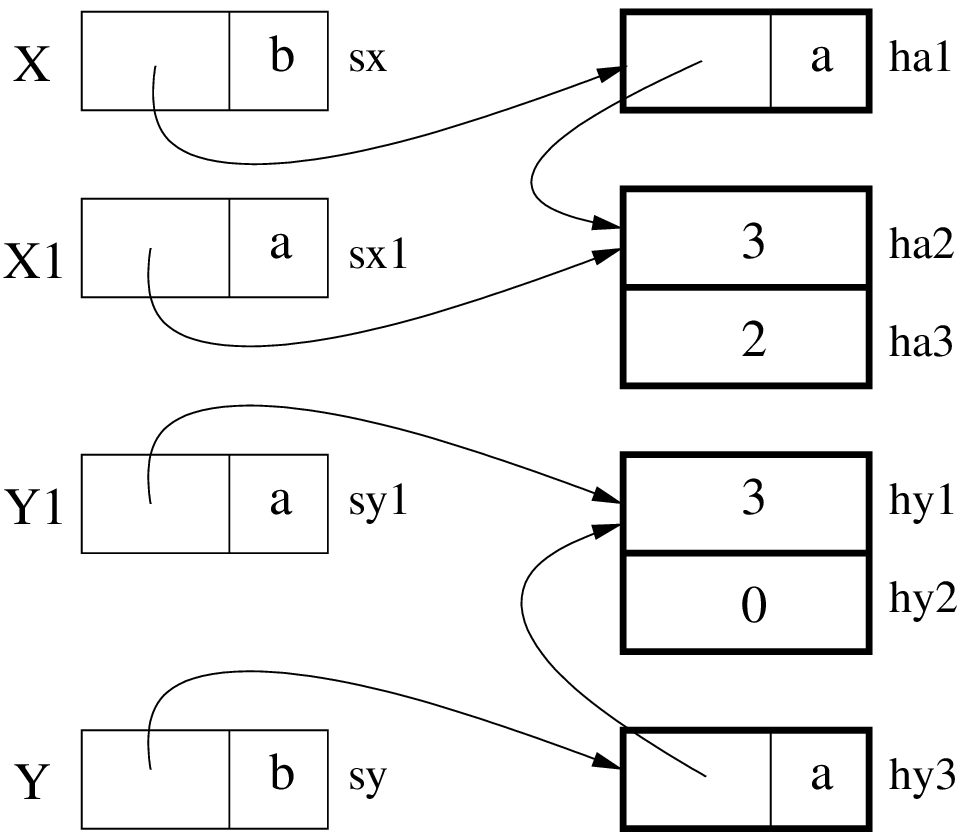}\par
\end{center}
\caption{No reuse}
\label{fig:no_reuse}
\end{minipage}
\hfill
\begin{minipage}[b][4.5cm][c]{6cm}
\begin{center}
\includegraphics[width=4.5cm]{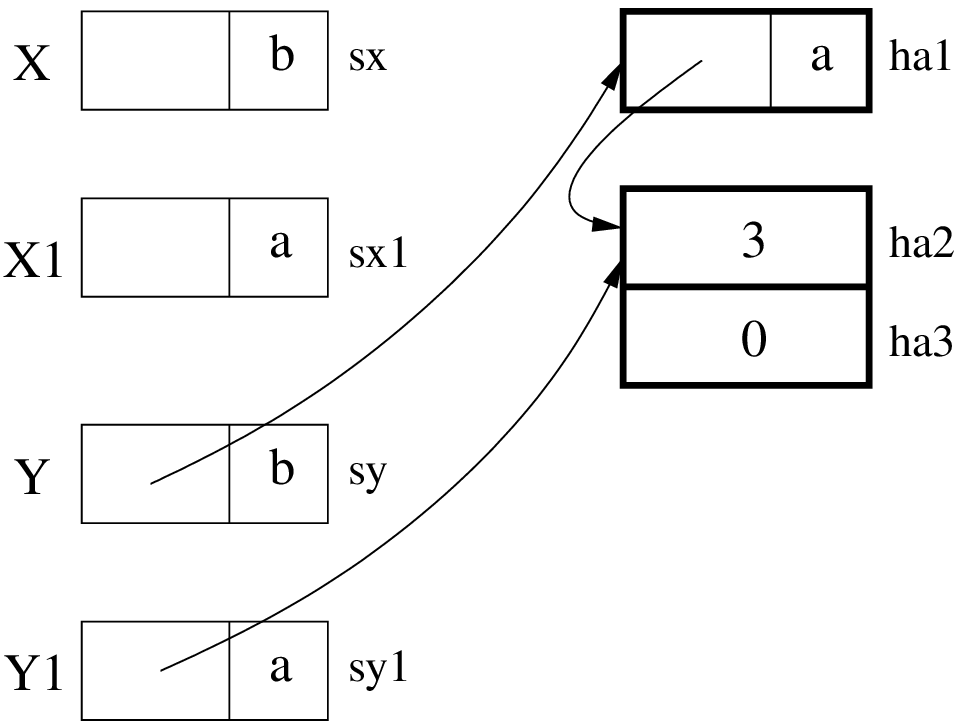}
\end{center}
\caption{Reuse}
\label{fig:reuse}
\end{minipage}
\end{minipage}
\end{figure}

Figure \ref{fig:no_reuse} shows the memory layout 
when calling {\tt convert1(A, B)} 
(Fig.\,\ref{fig:convert1}), 
where A is bound to {\tt b(a(3,east)} (Fig.\,\ref{fig:lowlevelX}). 

After deconstructing the input, new heap cells 
({\tt hy1}, {\tt hy2} and {\tt hy3}) are allocated
to create {\tt Y}, and the content of {\tt X} is partially copied into 
those cells. 
If it can be shown at compile-time that after this procedure call 
the term pointed at by {\tt X} will not be referenced during the
rest of the program (thus becoming available for reuse), 
then the deconstruction statements perform the last access ever to 
the concerned heap cells 
({\tt ha1}, {\tt ha2}, {\tt ha3}) after which they become 
garbage, and can 
be (re)used for {\tt Y} (Fig.\,\ref{fig:reuse}, the contents
of {\tt sx} and {\tt sx1} are no longer relevant).

The optimization could go further and detect that this reuse
only requires the update of one heap cell, namely {\tt ha3}. 
Yet currently we mainly focus on the memory usage of a program, 
execution time being only of indirect importance. Therefore
we do not try to optimize the number of field updates in the
presence of reuse. See also Section \ref{sec:improvements}.


\subsection{Types of Reuse}
\label{sec:typesreuse}

During reuse analysis we make a distinction between different kinds of 
reuses~\cite{31661}. 
The procedure shown in
Fig.\,\ref{fig:convert1} has {\em direct reuse} (of {\tt X}): it 
might contain the last reference to the cells of
{\tt X} which can then be reused by {\tt Y}. 
However, the reuse is {\em conditional}: if
the caller's environment does not correspond to the default
call pattern (e.g.\ keeping a reference to one of the cells
marked for reuse), reuse is not allowed. As this imposes very
harsh restrictions on the reuse possibilities, we introduced
the notion of {\em reuse conditions} which express the minimal conditions
a call pattern has to meet so that reuse is safe. These conditions
are expressed in terms of the variables involved (here {\tt X}). 
If the reuse is independent of the calling environment ({\tt X}
being a local variable), then we have {\em unconditional} reuse.

Given the reuse conditions, the next step of the reuse analysis is to
check for 
{\em indirect reuses}. Consider the following procedure: 
\begin{verbatim}
    :- pred generate(example). 
    :- mode generate(out) is semidet.
    generate(Y):- generate_2(X), convert1(X, Y).
\end{verbatim}
Assuming the default call pattern for {\tt generate}, 
the call to {\tt convert1} meets the condition
that after that call {\tt X} will not be used anymore. 
Hence, a reuse-version of {\tt convert1}
can be called and we say that {\tt generate} has 
{\em indirect reuse}. 
Moreover, {\tt X} is a local variable, so reusing it will always
be safe as it is independent of the call pattern
to {\tt generate}. This is an example of
{\em unconditional} (indirect) reuse. 
If {\tt X} would have been an input variable
to {\tt generate}, the indirect reuse would be conditional and
additional reuse conditions would be formulated.

\section{A Working Reuse Decision Approach}
\label{sec:selection}

Consider the predicate in Fig.\,\ref{fig:code} which converts a list
of data-elements into a new list of data-elements.  
While the data-flow analysis spots the datastructures that can 
potentially be reused, it is up to the reuse analysis to select those
reuses (direct and indirect) that yield the most interesting 
saving w.r.t.\ memory usage (and indirectly execution time). 


\begin{figure}[tb]
\begin{center}
\begin{minipage}[b]{10cm}
\begin{verbatim}
    :- type field1 ---> field1(int, int, int). 
    :- type field2 ---> field2(int, int). 
    :- type list(T) ---> [] ; [T | list(T)].
    :- pred convert2(list(field1), list(field2)).
    :- mode convert2(in, out) is det.

    convert2(List0, List):-
        ( % switch on List0
            List0 => [], List  <= []
        ;
            List0 => [Field1 | Rest0],      % (d1)
            Field1 => field1(A, B, _C),     % (d2)
            Field2 <= field2(A, B),         % (c1)
            convert2(Rest0, Rest),           
            List <= [Field2 | Rest]         % (c2)
        ).
\end{verbatim}
\end{minipage}
\end{center}
\caption{
Converting lists. 
}
\label{fig:code}
\end{figure}

\subsection{Deciding Direct Reuse}

A first restriction we impose is to limit reuses to
{\em local reuses}, 
i.e. a dead cell can only be reused in the same procedure
as where it is last accessed (deconstructed).
In Section \ref{sec:cellcache} 
we discuss techniques of how to lift this restriction. 
Furthermore, we consider that dead structures can only be reused by at
most one new structure.  Using the terminology of 
Debray~\cite{debray.iclp93}, we limit ourselves to the {\em simple} reuse
problem. It is not difficult to remove this limitation,
but it makes the reuse decisions more complex. 
We plan to lift this restriction in the future. 


%

The data-flow analysis of the example identifies the deconstructed
datastructures (at {\tt d1}, resp.\ {\tt d2}) 
as available for reuse. The procedure also contains
two constructions ({\tt c1} and {\tt c2}) where the memory
from the dead cells could be reused.

Each of the combinations yields an acceptable reuse-scheme. 
Yet, which one is the most interesting? 
It has been shown that this problem~\cite{debray.iclp93} can be
reformulated as an instance of the maximum weight matching problem for a
weighted bipartite graph.
However for simplicity of implementation
we have reduced this general matching problem to two
orthogonal decisions: imposing constraints on the allowed reuses,
and using simple strategies to select amongst different candidates for reuse. 
We will discuss each of these. 

\paragraph{Constraints on allowed reuses.} 

Constraints allow one to express common characteristics between the dead 
and the newly constructed cell and reflect the restrictions which can 
be imposed by the back-end to which a Mercury program is compiled. 

We have implemented the following constraints:
\begin{itemize}
\item Almost matching arities.
This constraint 
expresses the intuition that it can be worthwhile to reuse a dead cell,
even if not all memory-words are reused. 
This is indeed interesting if it can be guaranteed
that the superfluous words will be collected by the run-time garbage
collector within a reasonable delay. 
In our example, allowing a difference of size
one allows {\tt c1} and {\tt c2} to reuse the memory
available from either {\tt d1} or {\tt d2}. 
\item Matching arities. If the run-time system is not powerful 
enough to be used with {\em almost matching arities}, then a more
restrictive constraint can be used: only allow reuse between constructors
having the same arity. 
This means that in our example only {\tt d1} can be reused (by either
{\tt c1} or {\tt c2}). 
\item Label-preserving. 
Using the Java or .NET back-end, it is not
possible to change the type of run-time objects, therefore
reuse is only allowed if the dead and new cell have the
same constructor (label). 
For the example this means that the cell from {\tt d1} can
only be reused in {\tt c2}.
\end{itemize}

\paragraph{Selection strategies.}

When a cell can reuse different dead cells, a choice has
to be made (e.g.\ {\tt c1} can either reuse the cell available from 
{\tt d1} or {\tt d2}). Some choices yield better results than others.
We have experimented with two simple strategies: 
\begin{itemize}
\item Lifo. 
Traverse the body of the procedure and assign the reuses using a
last-in-first-out selection strategy. 
This means that when a choice is left for a given construction, choose
the cell which died most recently. 
The intuition is that after deconstructing a variable, it is very likely that
a new similar cell will be constructed in the same context. 

e.g. If {\tt c1} is allowed to reuse the cells from {\tt d1} or {\tt d2}, 
then according to this strategy, {\tt Field1} will be reused for constructing
{\tt Field2} and {\tt List0} for {\tt List}. 

\item Random. 
The intuition behind the lifo-strategy might not always be true, for
example in the presence of a disjunction\footnote{e.g. 
{\tt X => f(..), ( ... Y <= f(..) ; ... ), Z <= f(..)}:
as the first branch of the disjunction might not always be executed, it is
more interesting to allow {\tt Z} to reuse {\tt X} than {\tt Y}. }.
Therefore we
have added a simple strategy which randomly selects the dead cell
amongst all the candidates. 
\end{itemize}



\subsection{Deciding Indirect Reuse}

In order to decide whether a call to a procedure can be substituted by a call
to a reuse version of that procedure, we must be sure that such 
substitution is safe. 
This is tested by checking the reuse-conditions 
(under the assumption of a default call pattern).
If it is safe to call the reuse-version we have to decide
whether we will do so or not.

Here we have decided for simplicity by always calling the
reuse-version of a procedure if it is safe to do so. 
In Section \ref{sec:improvements} we discuss the drawbacks and suggest
a possible better solution. 

Suppose that for our previous example we would
only allow the reuse of the list-cells ({\tt d1} by {\tt c2}). 
Such reuse is conditional: the list
cell only dies iff it is not needed within the caller's context.
This condition has to be checked
for the recursive call. Under the default call pattern (see 
Section \ref{section:reuse}) {\tt Rest0} is dead at the moment of
the recursive call, hence the condition is satisfied, and the recursive
call can safely be substituted by a call to its reuse 
version\footnote{Note that this indirect reuse is in itself conditional: 
it can only be allowed if the list-cells of {\tt Rest0}
are not needed in the callers context.}.



\subsection{Splitting into Different Versions}

Once the possible direct and indirect reuses have been decided, there is
one remaining decision left: how many versions of a given procedure should 
be created? In our example, we might have detected three reuses: 
{\tt List0} reused by {\tt List}, {\tt Field1} by {\tt Field2}, 
and the indirect reuse (the recursive
call to the reuse version). 
We can generate 4 interesting versions of the initial procedure: 
a version with no reuse, a version reusing only {\tt List0}, a 
version reusing {\tt Field1} and a version reusing both
(where the reuse versions also include the recursive reuse call).
In general, for a procedure with
$n$ possible direct reuses, $2^n$ interesting versions can be
created. 
In our implementation we limit the number of versions 
to at most two: a version which imposes no conditions on the caller
(containing all possible unconditional reuses), and a version containing
all detected reuses.
In Section~\ref{sec:improvements} we briefly discuss other possibilities.


\section{Low Level Additions}
\label{sec:additions}

Given the previous decisions, a first CTGC-system
was implemented. 
Although good results were obtained for small programs
(e.g. naive-reverse), we ran into problems when analysing large
ones: 
\begin{itemize}
\item imprecision in the alias analysis had the effect that relatively
few cells were recognized as dead.
\item the number of aliases collected within a procedure 
became huge. This slowed down the operations manipulating them and the
CTGC process became too time consuming. 
\end{itemize}

\subsection{Enhancing the Aliasing Precision}

The underlying analysis for deriving alias-information uses the concept of
{\tt top} which expresses that all data parts might be aliased. This is
a safe abstraction in the case of total lack of knowledge about the 
possible existing
aliases at some program point. Once generated, this lack of information 
propagates rapidly as all primitive operations manipulating
it yield {\tt top} as well. 

Such a {\tt top} is generated in the presence of language constructs with
which the analysis cannot cope yet. These are procedures defined in terms of
foreign code (c, C++), higher-order calls and typeclasses. It is also
generated for procedures which are defined in other modules that have not
yet been analysed and for which no interface files have been generated yet. 

To obtain a usable CTGC-system, techniques were needed to limit the 
creation and propagation of {\tt top}. In our 
implementation, three techniques are used: 

\begin{enumerate}

\item {\em Using heuristics}. 
Based on the type- and mode- declaration of a procedure, one
can derive whether it can create additional aliases or not, without
looking at the procedure's body. This is the case when a procedure
uses unique objects (declared {\tt di} or {\tt uo}~\cite{mercury.ref}), 
or only has unique output variables\footnote{A procedure call cannot create
additional aliases between input variables as they must be ground at the
moment when the procedure is called.} or when the
non-unique output arguments are of a type for which 
sharing is not possible 
(integers, enums, chars, etc.). In all these cases, it is safe to conclude that
the procedure will not introduce new aliases. 

\item {\em Manual aliasing annotation for foreign code}. 
Important parts of the
Mercury standard library consist of procedures which are defined in terms of
foreign code. 
With the intention to be used mainly in this
standard library, we have extended the Mercury language such that foreign
code can be manually annotated with aliasing-information. 

\item {\em Manual iteration for mutual dependent modules}. 
The current com\-pi\-la\-tion-scheme of Mercury is not
yet able to cope with mutual dependent modules. 
Consider a module A in which some procedures are expressed in 
terms of procedures declared in a module B, and 
vice versa.
The normal compilation scheme is to compile one of the files, and then the
other one. 
In the presence of an optimizing compiler this is not enough. 
At the moment the first module is compiled, nothing is known from the second
one, yielding bad precision for the first one. 
This bad precision will propagate further to the second file as the second
file relies on the first one. 
Bueno et al.~\cite{ciao_lopstr} propose a new compilation scheme which is
able to handle these cases. As this requires quite some work, we make a work
around by allowing manually controlled incremental compilation. 
\end{enumerate}

\subsection{Making Compilation Faster: Widening the Aliasing}

While it is interesting to have more precise aliasing information than simply
{\tt top}, having more aliases also slows down the system. 
Now one can argue that speed is
not a major requirement of a CTGC-system as it is primarely intended to be used 
only at the final compilation phase of a program, but even for our
benchmarks we were not ready to wait hours for a module to compile.
Therefore, in order to produce a usable CTGC-system we have added a 
widening operator~\cite{cousot92comparing} which acts upon the aliases 
produced\footnote{This widening operator can be enabled on a per-module 
base. The user can also specify the threshold at which widening
should be performed: e.g. only widen if the size of the set of
aliases exceeds 1000.}. 

During the data-flow analysis, a datastructure is represented by its
full path down the term it is part of. Such a path is a concatenation of
selectors which selects the functor and the exact argument position in the 
functor\footnote{Infinite paths are avoided by simplifying full type trees
to type graphs. This is beyond the scope of this paper.}.
Aliases are expressed as pairs of datastructures. 

To illustrate this, let us consider the following definition of a tree type:
\begin{verbatim}
:- type tree ---> e ; two(int,tree,tree)
                  ; three(int,int,tree,tree,tree).
\end{verbatim}
After the construction {\tt V <= three(2,3,two(0,e,e),A,A)} 
(where {\tt A} is a
variable bound to another tree-term), 
the path $(three,3)\cdot(two,1)$ selects in {\tt V} the zero-integer. 
The path
$(three,3)$ selects in {\tt V} the whole datastructure corresponding to the
first subtree (namely {\tt two(0,e,e)}). 
In {\tt V}, the positions corresponding with the 
paths $(three,4)$ and $(three,5)$ are aliased.

For the aliasing information, we introduced {\em type widening} 
that consists of 
replacing a full path of normal selectors
by one selector, a so called {\em type selector}. The meaning of
a {\em type selector} is as follows: instead of selecting
one specific subterm of a term, it will select all the subterms which 
have the type expressed by the selector. In our example, 
the paths $(three,1)$, $(three,2)$, and \mbox{$(three,3)\cdot(two,1)$} all select
integer elements of {\tt V}. 
With type widening, all these selectors
are reduced to the selector $(int)$, i.e. the type of the
subterms which they select. 
The alias in {\tt V} (ie. between $(three,4)$ and $(three,5)$)
becomes
an alias between $(tree)$ and $(tree)$, hence expressing that 
all subtrees of {\tt V} might be aliased. If other
aliases between subtrees of {\tt V} exist, then they will all be replaced by
this one single alias, hence making the overall size of the set
of aliases smaller.

This widening leads to a considerable speed-up of the CTGC-system (compilation
of some modules taking almost one hour was now reduced to less than a minute).
Our results suggest that the overall precision remains sufficient in order
to detect the expected reuses for our benchmarks. 


\section{First Results}
\label{sec:results}

We have evaluated the effectiveness of our CTGC-system by comparing
memory usage and measuring compilation times. 
We have used toy benchmarks and one
real-life program. 
All the experiments were run on an Intel-Pentium III (600Mhz) with
256MB RAM, using Debian Linux 2.3.99, under a usual workload. The CTGC-system
was integrated into version 0.9.1 of the MMC. The reported memory information
is obtained using the MMC memory profiler. This profiler counts the total
number of memory words that are allocated on the heap\footnote{Note that
this count is independent of any run-time garbage collection.}. The timings
are averages of 10 runs each time. 
All the benchmarks are compiled using a non-optimized Mercury standard library
w.r.t.\ memory usage (hence no reuse in the library 
predicates\footnote{Normally, a Mercury system with CTGC would also have
the library modules compiled with CTGC in the same way as user modules.}). 
This allows
us to focus on the reuse occurring in the actual code of the benchmarks. 

The toy benchmarks comprise {\em nrev} (naive reverse of a list of 3000
integers), {\em qsort} (quick sort of a sorted list of 10000 integers), 
and {\em argo\_cnters} (a benchmark counting various properties of a file,
also used in~\cite{31661}). Table\,\ref{table:small_results} shows
the results. These are independent of the CTGC configuration used, as they 
all yield the same results here. For each of the benchmarks
every possible reuse is detected, yielding the expected savings in
memory usage and execution time.


\begin{table}
\begin{center}
\begin{tabular}{|l||c|c|c||c|c|c|c|}
\hline
       & \multicolumn{3}{c||}{No Reuse} & \multicolumn{4}{c|}{Reuse} \\
\cline{2-8}
module & C (sec) & M (Word) & R (sec) & C (sec) & M (Word) & m (\%) & R (sec) \\
\hline
nrev   & 1.49  & 9M & 1.51  & 11.79 & 6k & -99.9  & 0.32  \\ 
qsort  & 1.40  & 50M & 36.63  & 11.29 & 20k & -99.9 & 27.22  \\
argo\_cnters & 4.53 & 3.00M & 0.35 & 16.38 & 2.60M  & -13.3 & 0.32 \\
\hline
\end{tabular}
\end{center}
\caption{Toy benchmarks. 
C = compilation time.
M = number of allocated words. 
R = execution time.
m = relative reduction in memory usage.}
\label{table:small_results}
\end{table}


Next to small benchmarks, we found it important to evaluate 
the system on a real-life program, where the different constraints
and strategies do make a difference. 
The program we used is a ray tracer program 
developed for the ICFP'2000 programming
contest~\cite{icfp_contest2000} where it ended up fourth. 
This program transforms a given scene description into a rendered image. 
It is a CPU- and memory-intensive process, and therefore an ideal
candidate for our CTGC-system to be tested on. A complete description of this
program can be found at~\cite{icfp_mercury_entry}.

The program consists of 20 modules (5700 lines of code), 
containing mostly deterministic predicates. All modules
could be compiled without widening, except for one: {\em peephole}. This
module manipulates complex constructors and generates up to 11K aliases.
Without type-widening, the compilation of {\em peephole} takes 160 minutes.
With type-widening (at 500 aliases), it only takes 40 seconds. 
The compilation of the whole program with CTGC (and widening)
takes 5 minutes, compared to 1 minute for a normal compilation. 
As some of these modules depend on each other, the technique of manually 
iterating the compilation was used to obtain better results. For this
benchmark, the compilation had to be repeated 3 times to reach
a fixpoint (for a total time of 15 minutes). 
Each time every module was recompiled. In a smart compilation 
environment, most of the recompilations could be avoided. 

To measure the effects of the different constraints and strategies we 
have compiled the ray tracer with different CTGC-configurations. 
%
%
The first row of Table~\ref{table:icfp_results} 
shows the number of memory words and the  execution time (in seconds)
needed to render a set of 27 different scene descriptions (ranging from
simple scenes, to more complex ones) using a version 
of the ray tracer without CTGC. Rows 1 to 9 show the
relative memory usage and execution time of ray tracers compiled
using different CTGC-configurations for the same set of 
scene descriptions:
\begin{itemize}
\item Using the matching arities ({\em match})
or label-preserving ({\em same cons}) constraints, 
up to 24\% memory can be saved
globally. For some scene descriptions, this can go up to 30\%. 
There is also a noticeable speedup (14\%).
\item Using almost matching arities within a distance
of one ({\em within 1}) or two ({\em within 2}), 
much less memory is saved (only 10\%) with hardly
any speedup. 
The bad memory usage is not surprising as none of the selection
strategies takes into account the correspondance of the arities
between a new cell and the available dead cells. 
The bad timings are also explicable: 
with non-matching arities, reuse leaves
space-leaks which cannot immediately be detected by the current run-time 
garbage collector, hence the garbage collector will be called
more often. Improvements to the garbage collector are required. 
\item Globally, using the {\em random} selection strategy yields slightly
worse results than {\em lifo}. For some scene descriptions though,
results are better, but without spectacular differences. 
\item Row 9 shows the results of a ray tracer compiled using a 
version of the Mercury standard library {\em with} CTGC. There
is hardly any difference with Row 1, where libraries were used 
without CTGC. This is due to the fact that the ray
tracer makes a limited use of these libraries. 
\end{itemize}

Finally, a version of the ray tracer was built without type-widening
({\em lifo} and {\em matching arities}). 
Compared to row 1 in Table~\ref{table:icfp_results}
the overall memory usage difference is less than 1\%. The execution times
are comparable. 

\begin{table}[bt]
\begin{center}
\begin{small}
\include{summary}
\end{small}
\end{center}
\caption{ICFP-ray tracer using different 
CTGC-configurations.}
\label{table:icfp_results}
\end{table}


\section{Non-local Reuse: Cell Cache}
\label{sec:cellcache}

Currently we have assumed that all dying datastructures must be reused 
locally, i.e. within the same procedure in which they die.  
Hence quite some interesting possibilities of reuse could be missed.


We see three ways to achieve non-local reuses as well. 
The first and the most difficult is to extend the data-flow analysis 
to handle non-local reuse. 
The analysis would have to propagate possible dead cells and thus become
quite complex. 
It would also require intensive changes in the internal calling convention 
of procedures within the MMC as the address of the
cells to be reused would have to be passed between procedures. 
The second approach is to combine reuse analysis with inlining in such
a way that the cell death and subsequent reuse end up in the same procedure.
The third approach, which is the one we implemented, is to 
{\em cache dead cells}.
Whenever a cell dies unconditionally 
and cannot be reused locally, we mark it as {\em cacheable}. At runtime
the address of the cell as well as its size will be recorded in a cache
(or free list). 
Before each memory allocation the runtime system will first check 
the cell cache to see if a cell of the correct size is available and use
that cell instead of allocating a new cell.
This operation increases the time taken to allocate a memory cell in the
case of the cell cache being empty,
and hence should only be a win if the cell cache occupancy rate is
high.
It also avoids new
allocations so the overall cost of the runtime garbage collection system should
go down due to smaller heap sizes and less frequent need for garbage
collection. 

The {\em cc}-entries of Table~\ref{table:icfp_results} (Rows 10-17) 
show the results
of CTGC-confi\-gu\-ra\-tions combined with
the cell cache technique. 
Compared to the basic CTGC-confi\-gu\-ra\-tions, cell caching always increases
memory savings, going up to 49\% (for some scenes even 70\%). 
In the case of label-preserving or matching arities constraints, execution
time drops slightly. On the other hand, using almost matching
arities combined with cell caching increases the execution time. 



\section{Further Improvements}
\label{sec:improvements}

In the near future, we intend to explore a number of improvements to our 
system. 
First, for some procedures, several possibilities of reuse
are discovered, each one imposing its own reuse conditions.
Taken together, these reuse conditions are too restrictive on the
caller, hence hardly any calling environment
is able to satisfy them,
and no reuse is performed at all. 
A {\em top-down call-dependent version
splitting pass} could aid in generating more useful reuse-versions of
procedures, and avoid the generation of the useless ones. 

A second problem is the too absorbant effect of the notion of {\tt top}
currently used in the alias information. Once {\tt top} is encountered, 
it propagates all throughout the remainder of the code. 
Instead of {\tt top}, we could use 
{\em topmost substitutions}~\cite{full-prolog-esop96}: 
e.g.\ generating all possible combinations of aliases between 
the arguments of a called predicate, based on the types of these
arguments, either explicitly or in a more compact form 
(using type-selectors or 
keeping sets of variables, stating that these variables
might be aliased to each other in any possible way). 

In this paper we mainly focussed on memory savings, 
reasoning that saving memory implies less garbage collection, hence
diminishes the execution time. If execution time is of primary
concern, than more sophisticated reuse strategies will 
be needed. In the near future we will adopt the use of 
weighted graphs~\cite{debray.iclp93} where
the weights can be adjusted for minimizing memory usage
or execution time (taking into account the fields
that do not need to be updated). 
We will also consider splitting dead cells and reusing them
for different new cells.

In~\cite{Gudjonsson:1999:CTM}
the focus on execution time is even greater, trying to discover
almost every field not requiring an update, going even beyond
the boundaries of single procedures. This is indeed important
in Prolog, where the determinism of procedures is not necessarely
known at analysis time, and where given the underlying data-flow
analysis, each cell update requires extra
care in the case the value has to be reset upon backtracking. 
In Mercury, where determinism {\em is} known at compilation time, 
and where the analysis explicitly takes into account backtracking, this
is not a major issue. Therefore, it is not our immediate intention
to try to avoid every possible cell update. 




\if{0}
\section{Related Work}
\label{sec:related}

In this paper we have mainly focussed on techniques of reuse-selection 
that were easy to implement in the existing compiler. 
This is an instance of the more general reuse problem, 
where dead cells are allowed to be split up in chunks and reused
by different new cells separately. This problem was first
formalized in~\cite{debray.iclp93}. It shows that finding
an optimal solution is NP-complete. Restricting
the problem to reuse where splitting up dead cells is not allowed 
reduces it to a polynomial problem. 
Another approach was taken in~\cite{Gudjonsson:1999:CTM}
which focusses on optimizing Prolog programs w.r.t.\ execution time mainly,
hence trying to avoid every useless field update. This requires
to look even beyond the boundaries of single procedures and makes
the problem more complex. The notion of reuse maps and
savings function is used to express and evaluate one-one mappings
between dead cells and their reuses. In practice, the authors use
heuristics limiting the extent of the problem, and compute
alll possible reuse-maps of which they will choose the best one
(using the savings function). 

\fi


\section{Conclusion}
\label{sec:conclusion}

This paper describes a complete working compile-time garbage collection system
for Mercury, a logic programming language with declarations. 
The system
consists of three passes: data-flow analysis, reuse decision, and low
level code generation. The data-flow analysis based on~\cite{31661} detects
which cells become available for reuse. This paper presents 
easy implementable
restrictions, 
constraints and strategies for selecting realistic reuses.
In order to obtain a workable CTGC-system, 
low level improvements were introduced. 

A major contribution of this work is the integration of the CTGC
system in the Melbourne Mercury Compiler and its evaluation. 
Some small 
benchmarks were used, but also one real-life complex program, a ray tracer. 
Average global memory savings of up to 49\% 
were obtained, with a speedup of up to 17\%. 
It would be interesting to compare these results with the 
total potential of reuse within the program. This total potential could
be approximated using the techniques used in our first
prototype~\cite{31661} to predict the amount of reuse. 

Beside the proposed improvements 
the system could also be adapted to handle higher order calls and
type classes properly (instead of generating {\tt top} aliasing, and not
allowing reuse). Yet given the fact that many higher order calls are
specialized away by the compiler, we currently do not believe that the
overhead needed to deal with these language constructs is worthwhile. 


{\small
\bibliography{abi}
\bibliographystyle{abbrv}
}
\end{document}

%% file: summary.tex

\begin{tabular}{|l|c|c|c||c|c|c|c|}
\hline
& \multicolumn{3}{c||}{Configuration} & \multicolumn{2}{c|}{Memory} & \multicolumn{2}{c|}{Time} \\
& \multicolumn{3}{c||}{} &  \multicolumn{1}{c}{(kWord)} &  \multicolumn{1}{c|}{(\%)} &  \multicolumn{1}{c}{(sec)}  &  \multicolumn{1}{c|}{(\%)} \\
\hline
0 & \multicolumn{3}{c||}{no CTGC} & 1024795.51 & - & 362.31 & - \\
\hline
1 & lifo & match &  & 776707.92 &-24.21& 311.85 &-13.93 \\
2 & lifo & same cons &  & 791742.06 &-22.74& 313.57 &-13.45 \\
3 & lifo & within 1 &  & 916642.90 &-10.55& 361.84 &-0.13 \\
4 & lifo & within 2 &  & 917847.97 &-10.44& 359.90 &-0.67 \\
5 & random & match &  & 780838.58 &-23.81& 310.75 &-14.23 \\
6 & random & same cons &  & 795872.67 &-22.34& 312.70 &-13.69 \\
7 & random & within 1 &  & 920764.26 &-10.15& 359.14 &-0.87 \\
8 & random & within 2 &  & 921969.35 &-10.03& 355.08 &-2.00 \\
9 & lifo & match &  libs  & 775607.04 &-24.32& 320.32 &-11.59 \\
\hline
10 & lifo & match &  cc  & 513901.37 &{\bf -49.85}& 301.66 &-16.74 \\
11 & lifo & same cons &  cc  & 542626.80 &{\bf -47.05}& 304.20 &-16.04 \\
12 & lifo & within 1 &  cc  & 845603.55 &-17.49& 375.79 &3.72 \\
13 & lifo & within 2 &  cc  & 864722.90 &-15.62& 370.49 &2.26 \\
14 & random & match &  cc  & 518032.04 &{\bf -49.45}& 299.45 &-17.35 \\
15 & random & same cons &  cc  & 546757.48 &{\bf -46.65}& 302.79 &-16.43 \\
16 & random & within 1 &  cc  & 849724.90 &-17.08& 363.90 &0.44 \\
17 & random & within 2 &  cc  & 868844.29 &-15.22& 391.68 &8.11 \\
\hline
\end{tabular}